\documentstyle[12pt]{article}
\begin{document}
\pagestyle{empty}
\begin{flushright}
MADPH--96--945\\
UCD--96--17\\
May 1996
\end{flushright}

\vspace*{2cm}

\begin{center}
{\Large \bf Signals for Double Parton Scattering at the Fermilab Tevatron}\\
\vspace*{0.5cm}
Manuel Drees$^a$ and Tao  Han$^b$ \\
\vspace*{0.5cm}
{\it $^a$ Physics Department, University of Wisconsin, 
1150 University Ave., Madison, WI 53706, USA}\\
{\it $^b$ Department of Physics, University of California, Davis, CA  
95616, USA}\\
\end{center}
\vspace*{1.5cm}

\begin{abstract}
Four double-parton scattering processes are examined at the Fermilab
Tevatron energy. With optimized kinematical cuts and realistic parton
level simulation for both signals and backgrounds, we find large
samples of four-jet and three-jet+one-photon events with signal to
background ratio being 20\%-30\%, and much cleaner signals from
two-jet+two-photon and two-jet+$e^+e^-$ final states. The last  
channel may provide the first unambiguous observation of multiple parton
interactions, even with the existing data sample accumulated by the
Tevatron collider experiments.
\end{abstract}

\vskip 0.5cm

\clearpage
\setcounter{page}{1}
\pagestyle{plain}

There are good reasons to believe that multiple partonic interactions,
where two or more {\em pairs} of partons scatter off each other, occur
in many, or even most, $p \bar{p}$ collisions at the Tevatron
($\sqrt{s}=1.8$ TeV). On the theoretical side, multiple partonic
interactions are an integral part of the eikonalized minijet model
\cite{1}, which attempts to describe the observed increase of the
total $p \bar{p}$ cross section with energy in terms of the rapidly
growing cross section for the production of (mini)jets with transverse
momentum $p_T \geq p_{T,{\rm min}} \simeq 2$ GeV. Sj\"ostrand and van
Zijl \cite{2} also pointed out that including multiple interactions in
the PYTHIA event generator greatly improves the description of the
``underlying event'' in $p \bar{p}$ collisions. A similar result was
found recently by the H1 collaboration \cite{3} in a study of $\gamma
p$ collisions.

However, hadronic event generators have many ingredients. This makes
it difficult to draw unambiguous conclusions from such studies. It is
therefore desirable to search for more direct evidence for multiple
partonic interactions, using final states that are amenable to a
perturbative treatment. Clearly the cross section will be largest if
only strong interactions are involved. The simplest signal of this
kind is the production of four high$-p_T$ jets in independent partonic
scatters within the same $p \bar{p}$ collision \cite{4} ($4
\rightarrow 4$ reactions). Since energy and momentum are assumed to  
be conserved independently in each partonic collision, the signal for
a $4 \rightarrow 4$ reaction is two pairs of jets with the members of
each pair having equal and opposite transverse momentum. Various
hadron collider experiments have searched for this signature. The AFS
collaboration at the CERN ISR reported \cite{5} a strong
signal. However, the exact matrix elements for the QCD background $2
\rightarrow 4$ processes were not used and the size of the signal
claimed was considerably larger than expected. The UA2 collaboration
at the CERN SppS collider saw a hint of a signal, but preferred to
only quote an upper bound \cite{6}. More recently, the CDF
collaboration at the Fermilab Tevatron found evidence at the
2.5$\sigma$ level that $4 \rightarrow 4$ processes contribute about
5\% to the production of four jets with $p_T \geq 25$ GeV \cite{6a}.

While final states consisting only of jets offer large cross sections,
they suffer from severe backgrounds. There are three possible ways to
group four jets into two pairs. Further, the experimental error on the
energy of jets with $p_T \simeq 20$ GeV is quite large. Hence even
four-jet events that result from $2 \rightarrow 4$ background
processes often contain two pairs of jets with transverse momenta that
are equal and opposite within the experimental errors. The study of
``cleaner'' final states has therefore been advocated: The production
of two pairs of leptons (double Drell--Yan production) has been
studied in refs.\cite{7}, the production of two $J/\psi$ mesons in
refs.\cite{8}, and the production of a $W$ boson and a pair of jets in
refs.\cite{9}.  However, in our opinion none of these processes is
ideally suited for studying multiple partonic interactions. Double
Drell--Yan production offers a very clean final state, but the cross
section at Tevatron energies is very small once simple acceptance cuts
have been applied. The cross section for double $J/\psi$ production is
quite uncertain, since it depends on several poorly known hadronic
matrix elements \cite{10}. Finally, $W+$jets events can only be
identified if the $W$ boson decays leptonically, which makes it
impossible to fully reconstruct the final state.

Here we study mixed strong and electroweak final states: 
\begin{itemize}
 \item three jets and an isolated photon  ($jjj\gamma$); 

 \item two jets and two isolated photons ($jj\gamma \gamma$);  

 \item two jets and an $e^+ e^-$ pair ($jjee$).
\end{itemize}
For comparison, we also include
\begin{itemize}
 \item   four-jet final states (denoted by 4-jet).
\end{itemize}
We try to be as close to experiment as possible within a parton level
calculation. To this end we not only apply acceptance cuts, but also
allow for finite energy resolution, and try to model transverse
momentum ``kicks'' due to initial and final state radiation. We find
that the $jjj\gamma$ final state offers an only slightly better signal
to background ratio than the 4-jet final state does; note that the
combinatorial background is the same in these two cases. This
combinatorical background does not exist for the $jj\gamma \gamma$ and
$jjee$ final states, which offer much better signal to noise ratios,
at the price of small cross sections.

The calculation of our signal cross sections is based on the standard
assumption \cite{1,2,7,8,9} that the two partonic interactions occur
{\em independently} of each other. The cross section for a $4  
\rightarrow 4$ process is then simply proportional to the square of 
the $2 \rightarrow 2$ cross section:
\begin{equation} \label{e1}
\sigma ( 4 \rightarrow 4) = \left[ \sigma (2 \rightarrow 2) \right]^2  
/ \sigma_0.
\end{equation}
This assumption cannot be entirely correct, since energy--momentum
conservation restricts the available range of Bjorken$-x$ values of
the second interaction, depending on the $x$ values of the first
one. We include this (small) effect using the prescription of
ref.\cite{2}. In the eikonalized minijet model \cite{1} $\sigma_0$ is
related to the transverse distribution of partons in the
proton. Unfortunately total cross section data do not allow to
determine this quantity very precisely. We find values between about
20 and 60 mb, depending on the choice of the numerous free parameters
of the model. The recent CDF study \cite{6a} found $\sigma_0 = 24.2
^{+21.4}_{-10.8}$ mb, within the range that can be accommodated in
minijet models. We will take $\sigma_0 = 30$ mb in our numerical
analysis; the results can be scaled trivially to other values of
$\sigma_0$.

The for us relevant $2 \rightarrow 2$ cross section can be written as  
a sum of different terms:
\begin{equation} \label{e2}
\nonumber
\sigma( 2 \rightarrow 2 ) = \sigma (p \bar p \rightarrow j j X) +
 \sigma (p \bar p \rightarrow j \gamma  X)
+ \sigma (p \bar p \rightarrow \gamma \gamma X) 
+  \sigma (p \bar p \rightarrow e^+ e^- X),
\end{equation}
where $j$ stands for a high$-p_T$ jet. Inserting eq.(\ref{e2}) into
eq.(\ref{e1}) gives a $4 \rightarrow 4$ cross section that sums over
many different states; it should be obvious which terms in the sum are
of relevance to us. Note that this procedure gives an extra factor of
2 in the cross section for the production of final states made up from
two {\em different} $2 \rightarrow 2$ reactions (e.g, $jjj\gamma$)
compared to those produced from two identical reactions.  Partly for
this reason we only consider $j j \gamma \gamma$ configurations where
the two jets are produced in one partonic scatter and the two photons
in another. The other possible configuration ($j\gamma j \gamma$),
where each jet pairs up with one photon, also suffers from larger
backgrounds, since there are two ways to form such pairs. We use
leading order matrix elements in eq.(\ref{e2}), but we include the
contribution from $gg \rightarrow \gamma \gamma$, which enhances the
total $p \bar p \rightarrow \gamma \gamma X$ cross section by about
50\% at $\sqrt{s}=1.8$ TeV. We take MRSA' structure functions
\cite{12}; other modern parametrizations give very similar results. We
use the leading order expression for $\alpha_s$, with $\Lambda_{\rm
QCD} = 0.2$ GeV, and take the (average) partonic $p_T$ as
factorization and renormalization scale. We use exact leading order
matrix elements to compute the backgrounds from $2
\rightarrow 4$ processes. These have been computed in ref.\cite{13}
for the 4-jet final state, in ref.\cite{14} for the $jjj\gamma$ final
state, in ref.\cite{15} for $jj \gamma \gamma$ production, and in
ref.\cite{16} for $jjee$ production.

In order to approximately mimic the acceptance of the CDF and D0
detectors, we require all jets to have rapidity $|y_{\rm jet}| \leq
3.5$, while we require $|y_{e,\gamma}| \leq 2.5$ for electrons and
photons. We also require the isolation cut $\Delta R_{ij} \equiv
\sqrt{ \left( y_i - y_j \right)^2 + \left( \phi_i - \phi_j \right)^2} 
\geq 0.7$ for all combinations $ij$ of final state particles. We 
generally find that the $4 \rightarrow 4$ signal decreases more
quickly than the $2 \rightarrow 4$ background when the (transverse)
momentum of the outgoing particles is increased. The reason is that
the signal cross section contains four factors of parton densities,
while the background only has two. We therefore try to keep the
minimal acceptable $p_T$ as small as possible, subject to the
constraint that the event can still be triggered on. Specifically, we
chose
\begin{itemize}
\item[{\it i)}] for 4-jet: $p_T(j_1,j_2) \geq 20$ GeV, $p_T(j_3,j_4)  
\geq 10$ GeV 
\item[{\it ii)}] for $jjj\gamma$:
$p_T(\gamma, j_1) \geq 15$ GeV, $p_T(j_2,j_3) \geq 10$ GeV;
\item [{\it iii)}] for $jj\gamma\gamma$:
$p_T(\gamma_1, \gamma_2, j_1, j_2) \geq 10$ GeV;
\item[{\it iv)}] for $jjee$: $p_T(e_1, e_2) \geq 15$ GeV, $p_T(j_1,  
j_2) \geq 10$ GeV. 
\end{itemize}

The signal and background cross sections with only these basic
acceptance cuts included are listed in column 2 of Table 1 for the
4-jet and $jjj\gamma$ final states, and Table 2 for the $jj \gamma
\gamma$ and $jjee$ final states. We see that without further cuts, $4
\rightarrow 4$ processes only contribute between 7\% (4-jet) and 18\%
($jjee$), so additional cuts are clearly needed to extract the
signal. As expected from our previous discussion, the signal to
background ratio is worst for the 4-jet final state.

As mentioned earlier, in $4 \rightarrow 4$ processes two pairs of
particles are produced with equal and opposite transverse momenta,
$\vec{p_T}(1) = - \vec{p_T}(2)$ and $\vec{p_T}(3) = -
\vec{p_T}(4)$. However, additional radiation can change the kinematics
significantly, and the finite resolution of real detectors means that
we can require momenta to be equal only within the experimental
uncertainty.

In the presence of initial or final state radiation the transverse
momenta within a pair no longer balance exactly even if the resolution
was perfect. We include this effect only for the signal, since in the
background the final state particles in any case only pair up
``accidentally''; we do therefore not expect large effects on the
backgrounds.  We randomly generate transverse ``kicks'' for each of
the $2 \rightarrow 2$ processes in the signal. We assume that the
direction of the kick is not correlated with the plane of the hard
scattering. The absolute values $q_T$ of these additional transverse
momenta are generated according to the distribution
%
%
\begin{equation} \label{e3}
f(q_T) \propto \exp \left[ - \left({q_0} /{q_T} \right)^{0.7} \right]  
/q_T^2,
\end{equation}
with $0 < q_T \leq q_{T,{\rm max}}$. This function describes the
transverse momentum distribution \cite{17} of $W$ bosons produced at
$\sqrt{s} = 1.8$ TeV quite well, with $q_0 = 9$ GeV. We adopt this
choice of $q_0$ for the $jjee$ final state, which is dominated by the
production of real $Z$ bosons, but use the smaller value $q_0 = 4.5$
GeV for the other final states, which are characterized by a smaller
momentum scale.  Finally, we take $q_{T,{\rm max}} = 8$ GeV as our
default value; this assumes that one can reliably veto against jets
with transverse momentum exceeding this value.

We simulate finite energy resolutions by fluctuating the energies of
all outgoing particles (keeping the 4--vectors light--like), using
Gaussian smearing functions. The width of the Gaussian is given by
\begin{equation} \label{esmear}
 \delta(E) = a\cdot \sqrt{E} \oplus b \cdot E, 
\end{equation}
where $\oplus$ stands for addition in quadrature and $E$ is in GeV. We take
\begin{equation} \label{e4}
a_{\rm jet} = 0.80, \ \ b_{\rm jet} = 0.05, \ \ a_{e,\gamma} = 0.20,  
\ \
b_{e,\gamma} = 0.01,
\end{equation}
which roughly corresponds to the performance of the CDF detector. We
do not fluctuate the directions of the outgoing particles in this
step. These are, however, affected by the transverse ``kicks''
mentioned earlier. For this reason, and in order to allow for an error
in the determination of jet axes, we apply a relatively mild cut on
the azimuthal opening angle of each pair:
\begin{equation} \label{e5}
\cos(\phi_i - \phi_j) \leq - 0.9.
\end{equation}
This allows an opening angle as small as 154$^{\circ}$. As emphasized
earlier, in $4 \rightarrow 4$ processes, the members of a pair should
also have equal absolute values of $p_T$. As our final cut, we
therefore require
\begin{equation} \label{e6}
\left| | \vec{p_T}(i) | - | \vec{p_T}(j) | \right| \leq c_{ij}
\sqrt{ \delta^2 [ |\vec{p_T}(i) ] + \delta^2 [ |\vec{p_T}(j)|] },
\end{equation}
with $\delta(|\vec{p_T}|) = a \cdot \sqrt{|\vec{p_T}|} \oplus b \cdot
|\vec{p_T}|$ as in eqs.(\ref{esmear}) and (\ref{e4}).

Our results for signal and background with these additional cuts
included are summarized in the Tables. For the 4-jet and $jjj\gamma$
final states (Table 1) we always take $c_{12} = c_{34} \equiv c$, but
we occasionally allow $c_{ee,\gamma\gamma} > c_{jj}$ in the $jjee$ and
$jj \gamma \gamma$ final states. The reason is that the cut (\ref{e6})
is much more severe for $e^+e^-$ and $\gamma \gamma$ pairs than for
jet pairs, due to the better resolution of electromagnetic
calorimeters, see eq.(\ref{e4}). Inclusion of the transverse ``kick''
therefore leads to a significant loss of signal if we take
$c_{ee,\gamma\gamma} = 1$.  Although the stronger cut still gives a
slightly better signal to noise ratio, given the limited available
event sample employing a looser cut might give a statistically more
significant signal. We do not attempt to quantify this statement here,
since we have not included any reconstruction efficiencies in our
calculation. Finally, in the last three columns of Table 1 we increase
the cut on $\Delta R_{ij}$ from 0.7 to 1.2. This enhances the signal
to background ratio by about 20 to 25\%.

Switching on energy smearing and transverse momentum kicks, and
imposing the cuts (\ref{e5}) and (\ref{e6}) with $c=5$, reduces the
signal by typically a factor of 2. This reduction is almost entirely
due to the energy smearing.  Ignoring the transverse kicks for the
moment, in the signal both members of a pair have equal $|p_T|$. If it
falls below the cut--off value, {\em both} energies have to fluctuate
upwards for the event to be accepted. In contrast, the downwards
fluctuation of {\em one} energy can be sufficient to remove an event
from the sample. The reduction is smaller for $jjee$ production since
most electrons have typically $p_T \simeq M_Z/2$, well above the lower
limit. Fortunately the background is reduced even more in this step,
by a factor of 4 for 4-jet and $jjj\gamma$ and 9 for $jjee$ and
$jj\gamma\gamma$ final states, mainly due to the cut (\ref{e5}).
Making the cut (\ref{e6}) stricter, i.e. decreasing $c$, only slightly
enhances the signal to background ratio in Table 1. This is partly due
to the transverse kicks. Without them, the $jjj\gamma$ signal for
$c=1.0$ would be about 50\% larger. This indicates that restricting
additional jet activity as much as possible is quite important.

Although in Table 1 the optimized S/B ratios are only about 0.23 for
4-jet and 0.31 for $jjj\gamma$, the signals are statistically quite
significant; recall that the CDF and D0 experiments together have
accumulated about 200 pb$^{-1}$ of data.  We do not attempt to further
optimize the S/B ratio for these two processes because we do not trust
our parton level analysis, with a simplified treatment of finite
detector resolutions and the effect of parton showering, sufficiently
to extrapolate into the tails of distributions. Nevertheless, given
the normalization uncertainties of leading order QCD predictions, one
will have to study the shapes of various distributions, such as the
opening angle $\cos\phi_{ij}$, $\Delta R_{ij}$ and $p_T$ balancing
etc.  in order to convince oneself that a signal is indeed present.
Clearly the S/B ratio is much more favorable for the $jjee$ and
$jj\gamma\gamma$ final states (Table 2). For these final states
reducing $c$ from its starting point $c=5$ does increase this ratio
significantly. Recall that for a fixed value of $c$ the cut (\ref{e6})
is much more restrictive for $e^+e^-$ and $\gamma\gamma$ pairs than
for $jj$ pairs; this reduces the background more than the signal.  On
the other hand, this also has the effect that after imposing the cut
(\ref{e6}) with $c_{jj} = c_{\gamma\gamma} = 1$, the size of the
$jj\gamma\gamma$ signal depends quite sensitively on the treatment of
the transverse kick. Had we used $q_0 = 9$ GeV in eq.(\ref{e3}), as
appropriate for $W$ production, the signal would have been reduced by
a factor of about 0.7, while without any transverse kick it would have
been larger by a factor 1.6. Clearly this uncertainty can be reduced
by using the actual measured $p_T$ distribution of $\gamma \gamma$
pairs produced at the Tevatron. Fortunately the $jjee$ signal is less
sensitive to the ``kick'', since the electrons are usually so hard
that adding or subtracting a few GeV does not matter very much.  This
final state therefore offers our most promising and robust signal.

In summary, we have studied four different final states with a view of
establishing an unambiguous signal for multiple partonic interactions
in $p \bar p$ collisions at the Tevatron. The 4-jet and $jjj\gamma$
final states offer very large event samples, but with a S/B ratio
about 0.2 - 0.3. One must study the shapes of various kinematical
distributions for confirmation of the existence of the signal, as was
indeed done by the CDF collaboration in their study of the 4-jet final
state \cite{6a}. The situation is much more favorable for the
$jj\gamma\gamma$ and, especially, $jjee$ final states; in the latter
case one can increase the event sample by including muon pairs as
well. Although even in these channels the signal to noise ratio is
less favorable than what we found for four-jet production in $\gamma
\gamma$ collisions \cite{18}, a clear signal should be visible  
already in the present data sample.

Once a signal is found, it would be important to establish if the
normalization $\sigma_0$ in eq.(\ref{e1}) is indeed the same for
different processes, and independent of the Bjorken$-x$ range probed,
as assumed in minijet models. Further, it would be very interesting to
reduce the $p_T$ cut for at least some of the jets as much as
possible, so that one can get closer to the actual minijet
region. This could greatly enhance our understanding of ``minimum
bias'' physics, and give us some confidence that we can trust
extrapolations to LHC energies, where the understanding of overlapping
minimum bias events becomes a crucial issue in the assessment of the
viability of various ``new physics'' signals. Finally, such studies
might shed new light on the thirty-year old problem of the rising
total hadronic cross sections.

\subsection*{Acknowledgements}
We thank Walter Giele for sending us a computer code based on the
results of ref.\cite{13}. We also thank J. Huston for information on
experimental capabilities, and H. Baer for discussions of the
transverse kick. The work of M.D. was supported in part by the
U.S. Department of Energy under grant No. DE-FG02-95ER40896, by the
Wisconsin Research Committee with funds granted by the Wisconsin
Alumni Research Foundation, as well as by a grant from the Deutsche
Forschungsgemeinschaft under the Heisenberg program. T.H. was
supported in part by DOE under grant DE--FG03--91ER40674.


\noindent

\clearpage
\noindent
{\bf Table 1:} Signal and background cross sections, as well as their
ratios (S/B), for 4-jet production (in nb) and $jjj\gamma$ production
(in pb) at the Tevatron. In the first column only the basic acceptance
cuts on the transverse momenta, rapidities and on $\Delta R_{ij}$ have
been applied. In the second column we in addition apply the cuts
(\ref{e5}) and (\ref{e6}), with $c=5$. In the last three columns we
sharpen the $\Delta R$ cut to $\Delta R_{ij} \geq 1.2$, and gradually
reduce $c$ as indicated. Note that the ``basic'' cross sections have
been computed ignoring finite energy resolution and transverse
``kicks''; these effects have been included in the other columns, as
described in the text.

\begin{center}
\begin{tabular}{|c||c|c|c|c|c|}
\hline
& & $\Delta R_{ij}\geq 0.7$, & \multicolumn{3}{c|}{$\Delta R_{ij}\geq 1.2$} \\
& basic & $c=5$ & $c=5$ & $c=2$ & $c=1$
\\ \hline
$\sigma(4j)$(S) & 266 & 131  & 91  & 87   & 57  \\
$\sigma(4j)$(B) & 3,990 & 878  & 485  & 442  & 246  \\
S/B              & 0.067   & 0.15  & 0.19  & 0.20  & 0.23  \\
\hline
$\sigma(jjj\gamma)$(S) & 515  & 265  & 169 & 158 & 97 \\
$\sigma(jjj\gamma)$(B) & 5,370 & 1,310 & 611 & 571 & 311 \\
S/B         & 0.096  & 0.20 & 0.28 & 0.28 & 0.31 \\
\hline
\end{tabular}
\end{center}

\vspace*{1cm}
\noindent
{\bf Table 2:} Signal and background cross sections in pb, as well as
their ratios, for $jj\gamma\gamma$ production and $jje^+e^-$
production at the Tevatron. The notation is as in Table 1, except that
we use the basic isolation cut $\Delta R_{ij} \geq 0.7$ everywhere,
and allow different values for $c_1 \equiv c_{jj}$ and $c_2 \equiv
c_{ee}$ or $c_{\gamma \gamma}$.

\begin{center}
\begin{tabular}{|c||c|c|c|c|c|}
\hline
& basic & $c_1=c_2=5$ & $c_1=c_2=2$ & $c_1=1, \ c_2=2$ & $c_1=c_2=1$  
\\ 
\hline
$\sigma(jj\gamma\gamma)$(S) & 1.86 & 0.96 & 0.71 & 0.59 & 0.37 \\
$\sigma(jj\gamma\gamma)$(B) & 20.8 & 2.34 & 1.16 & 0.94  & 0.52  \\
S/B                         & 0.089 & 0.41 & 0.61 & 0.63 & 0.71 \\
\hline
$\sigma(jjee)$(S) & 3.45 & 2.01 & 1.42 & 1.07 & 0.62 \\
$\sigma(jjee)$(B) & 19.0 & 1.94 & 1.00 & 0.70 & 0.37 \\
S/B               & 0.18 & 1.04 & 1.42 & 1.53 & 1.68 \\
\hline
\end{tabular}
\end{center}

\end{document}